\documentclass{PoS}

\newcommand{\ba}{\begin{eqnarray}}
\newcommand{\ea}{\end{eqnarray}}
\newcommand{\be} {\begin{equation}}
\newcommand{\ee} {\end{equation}}

\title{Flavour physics from lattice QCD}

\ShortTitle{Flavour physics from lattice QCD}

\author{\speaker{Elvira G\'amiz}\thanks{The author's work is supported in part by the MICINN 
under Grants FPA2010-16696 and FPA2006-05294, and \emph{Ram\'on y Cajal} program; by Junta de 
Andaluc\'{\i}a under Grants FQM-101, FQM-330, and FQM-6552; and by European Commission under 
Grant No. PCIG10-GA-2011-303781. 
}\\
        CAFPE and Depto. de F\'{\i}sica Te\'orica y del Cosmos,
Universidad de Granada, E-18002, Granada, Spain\\
        E-mail: \email{megamiz@ugr.es}}

\abstract{I review recent results and prospects for improvement in calculations of
hadronic matrix elements relevant to quark flavour phenomenology. I will focus
on key quantities for CKM unitarity triangle analyses and for the study of
discrepancies between experiment and SM predictions for some flavour observables.
}

\FullConference{Xth Quark Confinement and the Hadron Spectrum,\\
		October 8-12, 2012\\
		TUM Campus Garching, Munich, Germany}

\begin{document}

\section{Introduction}
\label{sec:introduction}

\vspace*{-0.1cm}

The flavour physics program plays a dominant role in testing the Standard Model (SM) 
and searching for New Physics (NP), providing information which is complementary to 
direct searches in colliders. NP effects could be unveiled through the observation of 
deviations from the SM via high-precision measurements of low-energy observables in 
high-luminosity experiments. Indeed, there are several measurements of flavour 
observables for which there is a $2-3\sigma$ difference from SM predictions. These 
include $\sin(2\beta)$, the like-sign dimuon charge asymmetry, $B$ leptonic decays, 
unitarity triangle (UT) fits, and, more recently, the ratios of branching 
fractions of $B$ semileptonic decays to a $D$ and a $\tau$ over the corresponding 
decays to a $l=e,\mu$.

In order to test the Cabibbo--Kobayashi-Maskawa (CKM) description of the experimentally 
measured CP-violating and flavour changing processes, and investigate the origin of the 
discrepancies mentioned above, we need a determination
of the weak matrix elements involved in those processes with matching precision. 
In many cases, lattice QCD can provide those non-perturbative theoretical inputs 
from first principles and with errors at a few per-cent level. 
Accuracy in lattice calculations requires control over all the sources
of systematic error. In particular, it is essential to take into account
vacuum polarization effects in a realistic way, i.e., including up, down
and strange sea quarks on the gauge configurations' generation. The up and
down quarks are usually taken to be degenerate, so those simulations
are referred to as $N_f=2+1$. Two lattice collaborations (FNAL/MILC and ETMC) are 
now generating configurations which also include the effects of charm quarks 
on the sea, $N_f=2+1+1$. The first preliminary results for flavour quantities on those 
configurations are starting to appear. 

In the next Sections I will discuss the latest results for non-perturbative quantities 
relevant for flavour studies from lattice QCD calculations with all sources of 
systematic error addressed. 
Among other things, that means that I will focus on
simulations with $N_f=2+1$ sea quarks. 
Due to the increasing number of results available  
for some quantities from different collaborations working with $N_f=2+1$ configurations,   
it is important to perform averages of those results that can be used 
by non-experts. There are already two groups working on initiatives to compile results 
and provide averages: FLAG~\cite{FLAG} and Laiho-Lunghi-Van de Water~\cite{LLV}. 
A more ambitious project including members of those two averaging collaborations 
and new collaborators from Europe, Japan, and the US, 
the Flavour Lattice Averaging Group, FLAG-2, expects to provide its first set of 
averages soon~\cite{Colangelolat12}.

\vspace*{-0.1cm}

\section{Light quarks matrix elements}

\label{sec:light}

\vspace*{-0.1cm}

\subsection{$\vert V_{us}\vert$ from 
$K$ leptonic and semileptonic decays}

The lattice calculation of pseudoscalar decay constants,
together with experimental measurements of pseudoscalar
leptonic decay widths, can be used to extract the value
of the CKM matrix elements involved in those processes. 
The decay constants are easy to calculate on the lattice with high precision. 
The accuracy  achieved is even higher for ratios of decay constants, 
for example $f_K/f_\pi$, since 
many systematic uncertainties and statistical fluctuations cancel partially or 
completely between numerator and denominator. $f_K/f_\pi$ has been widely studied 
on the lattice and the precision has reached the subpercent level~\cite{fKoverfpi}. 
The average of the published results with $N_f=2+1$ is~\cite{LLV} 
$f_K/f_\pi=1.1936\pm0.0053$. 
Agreement between different collaborations also provides a good check of lattice 
methodologies. The phenomenological interest of this quantity stems from the fact that 
it is related to the ratio $\vert V_{us}\vert/\vert V_{ud}\vert$ and experimental 
measurements of $K$ and $\pi$ leptonic decay widths, so it can be used 
to extract a value for $\vert V_{us}\vert$. Using the experimental average in 
\cite{Antonelli10}, and the lattice average for $f_K/f_\pi$ above, we get 
$\vert V_{us}\vert=0.2252(11)$. 

Precise determinations of $\vert V_{us}\vert$ provide stringent tests of 
first-row unitarity and give information about the scale of NP~\cite{Cirigliano10}.
This CKM parameter can also be extracted from experimental data on $K$ 
semileptonic decay rates, given the value of the vector form factor 
at zero momentum transfer, $f_+^{K\pi}(0)$. The parameter $f_+^{K\pi}(0)$ is also 
one of the inputs used in UT analyses. The only unquenched lattice 
calculations available for this form factor~\footnote{After this conference,   
the FNAL/MILC collaboration has provide another result~\cite{KtopiFNALMILC12}, compatible 
with those by the RBC/UKQCD and ETMC collaborations, but with a larger central value and 
smaller errors. The FNAL/MILC result implies $|V_{us}| = 0.2238\pm0.0009\pm0.0005$.} 
are the ones by the RBC/UKQCD and ETMC collaborations~\cite{fplus}. 
The average of the two calculations is~\cite{LLV}  $f_+^{K\pi}(0)=0.9584(44)$, 
which gives $\vert V_{us}\vert=0.2257(11)$, in perfect
agreement with the value extracted from leptonic decays and also in agreement
with unitarity. 
Although these errors are already very small, there is a lot of room 
for improvement: more lattice spacings, more sophisticated extrapolation methods, 
physical light quark masses, $\dots$ 
The RBC/UKQCD, FNAL/MILC, and JLQCD collaborations have already presented 
preliminary results incorporating some of those improvements in~\cite{fKpiproceedings}.

\subsection{$K^0-\bar K^0$ mixing}

The information coming from neutral Kaon mixing provides one of the most stringent 
constraints in UT analyses. Until a few years ago, the limiting factor to 
exploit that constraint 
was the uncertainty in the calculation of the hadronic matrix element which encodes the 
non-perturbative physics of the process, parametrized by the bag parameter $B_K$. 
This is no longer true thanks to several $N_f=2+1$ lattice QCD calculations,  
four of them in the last year~\cite{BK2012}, 
which have reduced the error of $B_K$ under $2\%$. The average of all 
$N_f=2+1$ calculations is~\cite{LLV} $\hat B_K = 0.7643(97)$. 
The dominant errors in the $\vert\varepsilon_K\vert$ constraint in UT analyses are now the 
uncertainty in $\vert V_{cb}\vert$, which enters as the fourth power in the theoretical 
prediction for $\vert\varepsilon_K\vert$, and the error in the NNLO perturbative 
coefficient $\eta_{cc}$~\cite{etacc}. 
This year we have also had the first unquenched calculations of the matrix
elements of the four-fermion operators contributing to $K^0-\bar K^0$ in extensions of
the SM by the RBC/UKQCD Collaboration~\cite{Boyle:2012qb} ($N_f=2+1$ calculation
at a single lattice spacing) and by the ETMC Collaborarion~\cite{BKBSM} ($N_f=2$
calculation in the continuum limit).

\section{Heavy quark phenomenology}

\subsection{Heavy-light decay constants}

\label{sec:decayconst}

The decay constant of the $D_s$ meson, $f_{D_s}$, has generated a vivid interest in 
both experimentalists and lattice QCD phenomenologists in the last years, 
especially since 2007, when an important reduction in the errors of the lattice 
calculation of $f_{D_s}$ made possible to observe a $3.8\sigma$ disagreement between 
the lattice and experimental averages. Later improvements in both the theoretical and 
the experimental sides, however, shift both numbers and reduced the difference. 
Nevertheless, the most precise lattice calculations, including those in a preliminary 
stage, tend to give smaller values of $f_{D_s}$ than experiment. In particular, 
two $N_f=2+1+1$ preliminary results from the ETMC and the FNAL/MILC collaborations, 
are around $(245-250)~{\rm MeV}$~\cite{fDNf2+1+1}. 
The lattice averages, which only include the $N_f=2+1$ 
calculations already complete, are~\cite{LLV,Bazavov:2011aa,Na:2012iu}
\ba
\hspace*{-0.5cm}f_D = 209.2(3.9){\rm MeV}\,, \,\,\,
f_{D_s} = 248.6(3.0){\rm MeV}\,.\hspace*{-0.7cm}
\ea
The value of $f_D$ agrees with experiment~\cite{CLEOfD}, $f_D=206.7\pm8.9$, while 
$f_{D_s}$ is around $2\sigma$ lower than the experimental average~\cite{RosnerStone2012}, 
$f_{D_s}=(260.0\pm 5.4){\rm MeV}$. The preliminary BESIII result in~\cite{BES2012}, 
$f_D=(203.9\pm6.0){\rm MeV}$, agrees well with the CLEO 
result and the lattice average. 

UT fits are very sensitive to $f_B$ and different processes with potential to show up 
NP effects depend on $f_B$ or $f_{B_s}$, so any improvement in the decay
constants calculations as well as in the understanding of the 
$\vert V_{ub}\vert^{exc.}/\vert V_{ub}\vert^{inc.}$ disagreement is very important. 
There have been three lattice $N_f=2+1$ calculations of this parameter in the 
last two years by the HPQCD~\cite{HPQCDfB1,HPQCDfB2} and the 
FNAL/MILC~\cite{FMILCfB} collaborations, 
which have reduced the error at the $2.5\%$ level. The smallest error 
in \cite{HPQCDfB2} is achieved by determining the ratio $f_{B_s}/f_B$ using a 
non-relativistic description (NRQCD) of the $b$ quarks together with the $f_{B_s}$ 
determination in \cite{HPQCDfB1}, which employs relativistic $b$ quarks.  
By using the ratio $f_{B_s}/f_B$ the dominant systematic errors associated with the 
effective NRQCD description are partially cancelled, so this determination is nearly 
free of uncertainties due to an effective description of the $b-$quark.  
The average values of $f_B$ and $f_{B_s}$ from these three calculations are~\cite{LLV}
\ba\label{eq:fB}
\hspace*{-0.5cm}f_B = 190.6(4.7){\rm MeV}\,, \,\,\,
f_{B_s} = 227.6(5.0){\rm MeV}\,.\hspace*{-0.7cm}
\ea
The direct comparison of the results in Eq.~(\ref{eq:fB}) with experimental measurements
of the $B$ leptonic decay width is problematic due to the need of the value of the CKM
matrix element $\vert V_{ub}\vert$ 
(whose inclusive and exclusive determinations disagree at the $3\sigma$ level) and the
$\sim 2\sigma$ disagreement of BaBar~\cite{BaBarBtolnu} and Belle's~\cite{BelleBtolnu} 
measurements. Nevertheless, Belle new result seems to alleviate the tension between
theory and experiment previously observed.

\subsection{$B$ semileptonic decays}

\label{sec:Bsemileptonic}

There exist $\sim 2-3\sigma$ tensions between the inclusive and the exclusive
determinations of both $\vert V_{ub}\vert$ and $\vert V_{cb}\vert$. In addition to
experimental measurements on, for example, $B\to\pi l\nu$ and $B\to D(D^*) l\nu$,
respectively, the exclusive determination of those CKM elements need as input
form factors that can be calculated with high precision using lattice QCD techniques. 
There have not been new lattice QCD calculations of the form factors describing the
$B\to\pi l\nu$ decay since 2008~\cite{BtopiFMILC}, although several analyses are
in progress~\cite{Btopiproc}. Using the lattice results in Ref.~\cite{BtopiFMILC} and the
latest experimental data~\cite{HFAG12}, $\vert V_{ub}\vert_{{\rm exclusive}}
= (3.23\pm0.30)\cdot 10^{-3}$.

For the exclusive determination of $\vert V_{cb}\vert$, the state-of-the art
calculation of the relevant form factors is the FNAL/MILC analysis in~\cite{BtoDstar}, 
which studies the decay $B\to D^* l\nu$ at zero recoil.
The updated result is $\vert V_{cb}\vert =(39.7\pm0.7\pm0.7)\cdot 10^{-3}$, where
the first error is experimental and the second one the uncertainty in the lattice
calculation of the form factors. The FNAL/MILC collaboration is doing an extensive
study of both $B\to D^* l\nu$ and $B\to D l\nu$ decays at zero and non-zero
recoil~\cite{FMILCBtoD}, providing two independent modes for the extraction of
$\vert V_{cb}\vert$. This study will also provide checks of the shape of the
form factors, in addition to $\vert V_{cb}\vert$.

The FNAL/MILC collaboration recently analyzed a subset of their
$B\to D l\nu$ data to calculate the ratio of branching fractions 
$R(D)={\cal B}r(B\to D\tau\nu)/{\cal B}r(B\to D l \nu) =0.316(14)$~\cite{RDFNALMILC}. 
Their value is $\sim 1.7\sigma$ smaller than the recent experimental measurement by 
the BaBar collaboration~\cite{RDBaBar}. They found that the value of the ratio is 
very sensitive to differences in the scalar form factor, so one should be cautious 
in using indirect estimates of the form factors to constrain NP models in other 
decay channels such as $B\to D^* \tau\nu$~\cite{RDFNALMILC}. Given the present tensions, 
not only for $R(D)$ but for $R(D^*)$, and the possible indications of NP that could be 
extracted from a combined analysis of both set of decays, unquenched lattice QCD 
calculations of those two ratios should be a priority. Together with a determination of 
$\vert V_{cb}\vert$ and the shape of the form factors describing $B\to D(D^*) l\nu$ decays, 
the final FNAL/MILC analysis including the complete set of data will  
also provide an improved determination of $R(D)$ as well as a calculation of $R(D^*)$. 
Current experimental measurements of these 
ratios are statistics-limited, so Belle II  should significantly reduce the 
errors of those measurements. 
The Cambridge group and the FNAL/MILC collaboration have also presented preliminary 
results for the different form factors describing the rare decays 
$B\to K(K^*) l^+l^-$~\cite{raredecaysHPQCD,raredecaysFMILC}.

\subsection{$D$ semileptonic decays}

The semileptonic modes $D\to K l\nu$ and $D\to\pi l\nu$ allow us 
not only to extract the values of the CKM matrix elements $|V_{cs(d)}|$,
but to test lattice QCD techniques and methodology by comparing the shape
of the corresponding form factors, $f_+^{DK(\pi)}(q^2)$, with experimental data.
The method developed by the HPQCD collaboration that employs a Ward identity to relate 
the matrix element of a vector current to that of the corresponding scalar current,  
and the use of highly improved lattice discretizations to treat the charm quarks 
relativistically, have allowed a reduction of the errors in the lattice calculation of the form 
factors $f_+^{D\to\pi}(0)$ and $f_+^{D\to K}(0)$ from around a $10\%$ to a 
$5\%$ and a $2.5\%$ respectively. The average of the results 
in~\cite{HPQCDDtoKpi,FMILCDtoKpi} by 
the HPQCD and FNAL/MILC collaborations are 
\ba
\hspace*{-0.5cm}\vert V_{cs}\vert = 0.961(11)(24)\,,\,\,\,
\vert V_{cd}\vert =  0.225(6)(10)\,,\hspace*{-0.5cm}
\ea
where the first error is from experiment and the second one is the 
lattice error in $f_+^{D\to K(\pi)}(0)$. 
This is compatible with unitarity ($\vert V_{cs}\vert = 0.97345(16)$ and 
$\vert V_{cd}\vert = 0.2252(7)$), and it is competitive with the determination 
of $\vert V_{cd}\vert$ from neutrino scattering, 
$\vert V_{cd}\vert=0.230(11)$~\cite{PDG12}.

There are several on-going projects that study the dependence of the form factors
on the momentum transfer, showing a good description of the experimentally measured
shape.  The final goal is performing a global fit, including
all available experimental data on the decay rates and the lattice determination of
the form factors at several values of the momentum transfer, to extract the value of
the corresponding CKM matrix element in an optimal way as well as testing the
consistency of the lattice description throughout the energy range. HPQCD has quoted a 
preliminary number extracted from a global fit to Belle, BaBar, and CLEO data,  
$\vert V_{cs}\vert = 0.965(14)$~\cite{Koponencharm12}.

\subsection{Neutral $B-$meson mixing}

\label{sec:Bmixing}

It has been argued that differences observed between measurements of some flavor
observables and the corresponding Standard Model (SM) predictions may be due to beyond the
SM (BSM) physics affecting the neutral $B$-meson mixing processes~\cite{UTfits1,UTfits1bis}. 
Although the most recent analysis seem to indicate that there are not large BSM contribution 
to neutral $B$-meson mixing~\cite{NiersteCKM12}, the future will bring new twists, and 
precise calculations of the non-perturbative inputs parametrizing the mixing in the SM 
and beyond are necessary for a thorough understanding of quark flavor physics. 
The current status of $N_f=2+1$  lattice calculations of the non-perturbative quantities 
parametrizing the mass differences between the heavy and the light 
mass eigenstates in both the $B^0_d$ and $B^0_s$ systems, as well as the SU(3) breaking 
ratio $\xi$, is summarized in Tab.~\ref{tab:mixing}. The HPQCD~\cite{BmixingHPQCD} 
and FNAL/MILC~\cite{BmixingFMILC} collaborations 
use the same light quark formulation, but a different description for the $b$ quarks. 
In the exploratory study by the RBC/UKQCD collaboration in the table
heavy quarks are static \cite{BmixingRBC}. 
The average of the results in Tab.~\ref{tab:mixing} for $\xi$ 
gives the value $\xi=1.251\pm0.032$.

\begin{table}[t]
\begin{center}
\begin{tabular}{cccc}
\hline
\hline
 & HPQCD & FNAL/MILC & RBC/UKQCD \\
\hline
$\xi$ & 1.258(33) & 1.27(6) & 1.13(12) \\
\hline
$B_{B_s}/B_{B_d}$ & 1.05(7) & 1.06(11) & - \\
\multicolumn{4}{c}{HPQCD: $f_{B_s}\sqrt{\hat B_{B_s}} = 266(6)(17)~{\rm MeV}$,\hspace*{0.4cm}
$\hat B_{B_s}=1.33(6)$}\\
\multicolumn{4}{c}{HPQCD: $f_{B_d}\sqrt{\hat B_{B_d}} = 216(9)(13)~{\rm MeV}$,\hspace*{0.4cm}
$\hat B_{B_d}=1.26(11)$}\\
\hline
\hline
\end{tabular}

\end{center}

\vspace*{-0.3cm}
\caption{$B-$meson mixing parameters. $\xi$ is defined as the ratio of the
parameters in the second and third rows. In the case where there are two errors,
the first one is statistical and the second one systematic. \label{tab:mixing}}
\end{table}

Beyond the SM, the mixing parameters can have contributions from $\Delta B=2$ 
four-fermion operators which do not contribute in the SM. The matrix elements of 
the five operators in the complete basis describing 
$\Delta B=2$ processes, together with the Wilson coefficients for those operators 
calculated in a particular BSM theory and the experimental measurements of the mixing 
parameters, can provide very useful constraints on that BSM theory. There is not a final 
unquenched lattice calculation of the matrix elements of all the operators in the 
$\Delta B=2$ effective hamiltonian, but FNAL/MILC and ETMC presented preliminary 
results for the complete basis in~\cite{Bmixinglatt2011,MixinglatETMC12}. Those preliminary 
results include the matrix elements needed for the determination of
the decay width differences, $\Delta\Gamma_{d,s}$, for which there is not currently an 
unquenched calculation in the continuum limit.

The authors of Ref.~\cite{BurasBmumu} suggested that the branching fractions of the rare 
decays $B_q\to \mu^+\mu^-$ (for $q=s,d$) could be determined from the experimental 
measurement of the mass difference in the neutral $B_q$-meson system, $\Delta M_q$, and 
the lattice calculation of the bag parameter $\hat B_{B_q}$ since 

\vspace*{-0.5cm}
\ba\label{eq:BagBmumu}
{\cal B}r ( B_q \to \mu^+ \mu^- )/\Delta M_q
& = & [{\rm known\,factors}]/\hat B_q
\ea

\vspace*{-0.3cm}
In order to compare experimental measurements and the theory predictions for the decay 
rate of $B^0_s$, one should include the effects of a non-vanishing 
$\Delta\Gamma_s$~\cite{BsmumuFleischer}. This can be done in the SM by rescaling the 
theory prediction by $1/(1-y_s)$, where 
$y_s\equiv\tau_{B_s}\Delta\Gamma_s/2$~\cite{BsmumuFleischer}. 
Multiplying Eq.~(\ref{eq:BagBmumu}) by that factor for the $B^0_s\to\mu^+\mu^-$ decay 
and using the HPQCD determination of the bag 
parameters $\hat B_{B_s}=1.33(6)$ and 
$\hat B_{B_d}=1.26(11)$~\cite{BmixingHPQCD}; together with $\tau_{B_s}=1.497(15)ps$, 
$\tau_{B_d}=1.519(7)ps$~\cite{PDG12}, and  $\Delta\Gamma_s = 
0.116\pm0.019ps^{-1}$~\cite{LHCbDeltaGamma}, one gets 
\ba\label{eq:Btomumu}
{\cal B}r                                  
( B_s \to \mu^+ \mu^- )\vert_{y_s} =  (3.65\pm0.20)\times 10^{-9}\,,
\quad{\cal B}r ( B_d \to \mu^+ \mu^- ) =  (1.04\pm0.09)\times 10^{-10}\,.
\ea

The direct calculation of these branching fractions have become competitive with the one 
in (\ref{eq:Btomumu})~\cite{BurasGirrbach} thanks to the recent improvements in the 
calculation of the $B$-meson decay constants on the lattice described in 
Sec.~\ref{sec:decayconst}. 
Including the correction factor $1/(1-y_s)$ for the $B_s\to\mu^+\mu^-$ decay rate, 
and using the same inputs as in Ref.~\cite{BurasGirrbach} 
except for $f_B$ and $f_{B_s}$, for which I use the averages described in 
Sec.~\ref{sec:decayconst}, and $\tau_{B_s}$, which I take
equal to its PDG 2012 value, $\tau_{B_s}=1.497(15)ps$, I get
\ba\label{eq:Btomumudirect}
{\cal B}r ( B_s \to \mu^+ \mu^- )\vert_{y_s} =  (3.64\pm0.23)\cdot 10^{-9}\,,
\quad
{\cal B}r ( B_d \to \mu^+ \mu^- ) = (1.07\pm0.10)\cdot 10^{-10}\,.
\ea
The agreement between the two set of numbers in (\ref{eq:Btomumu}) and
(\ref{eq:Btomumudirect}) is excellent, giving us confidence on the SM prediction for
these branching fractions. This is very important since LHC is approaching the SM
predictions, with the first evidence for one of these two processes recently 
reported by LHCb~\cite{LHCb:2012ct}. The LHCb measurement~\cite{LHCb:2012ct},   
${\cal B}r ( B_s \to \mu^+ \mu^- )= \left(3.2^{+1.5}_{-1.2}
\right)\cdot 10^{-9}$, is consistent with the SM prediction in 
Eqs.~(\ref{eq:Btomumu}) and (\ref{eq:Btomumudirect}). 

Another recent contribution of lattice QCD to the study of $B_s \to \mu^+ \mu^-$ 
rare decays has been the calculation by the FNAL/MILC collaboration 
of form-factors ratios between the semileptonic decays $\bar B^0 \to D^+l^-\bar\nu$ and 
$\bar B^0_s \to D^+_sl^-\bar\nu$~\cite{fragmfrac}. These ratios are a key theoretical 
input in a new strategy to determine the fragmentation fractions of neutral $B$ 
decays, which are needed for the experimental measurement of 
${\cal B}r ( B_s \to \mu^+ \mu^- )$. The result for the ratio of form factors, 
$f_0^s(m_\pi^2)/f_0^d(m_K^2)=1.046(44)(15)$, gives a value of the fragmentation fraction 
which agrees well with that of the 
$D_s^+\pi^-/D^+K^-$ hadronic method and with LHCb's determination via a method 
employing semileptonic decays. 
This calculation included only a subset of the $B\to D l\nu$ data which 
is being analyzed by the FNAL/MILC collaboration for the determination of 
$\vert V_{cb}\vert$, so we can expect a considerable improvement when 
they finish analyzing their full data set.

\section{Outlook}

Lattice QCD calculations of non-perturbative parameters relevant for flavour phenomenology 
have achieved accuracies at the per-cent level for many key quantities. 
The agreement between results from different collaborations 
for the same quantities also allows an important check of lattice methods, especially in
the light and charm sectors. In the bottom sector, the first $N_f=2+1$ results using a 
relativistic approach are very promising. For the next two years we expect new results 
from the FNAL/MILC, ETM, and RBC/UKQCD collaborations for decay constant, $B^0-\bar B^0$ 
mixing, and $B\to\pi l\nu$. 
One of the important steps in the improvement of lattice calculations expected in 
the next year is using simulations at the physical light quark masses. This would 
drastically reduce one of the main systematic, the one associated with the chiral 
extrapolation. 
There will be results with physical quark masses within a year from 
different collaborations: FNAL/MILC, BMW, RBC/UKQCD, and PAC-CS. 
The reduction of the dominant sources of errors will make necessary to 
include some uncertainties which were subdominant until now like isospin breaking, 
electromagnetic effects, or dynamical charm quarks. 
Another target for lattice calculations is going beyond gold-platted 
quantities and develop methods to calculate more demanding quantities 
($K\to\pi\pi$, weak decay to resonances, non-local operators, ...).


\begin{thebibliography}{99}
\bibitem{FLAG}
 G.~Colangelo {\it et al.},
  Eur.\ Phys.\ J.\ C {\bf 71} (2011) 1695
  [arXiv:1011.4408 [hep-lat]].

\bibitem{LLV}
J.~Laiho, E.~Lunghi and R.~S.~Van de Water,
  Phys.\ Rev.\ D {\bf 81} (2010) 034503
  [arXiv:0910.2928 [hep-ph]]. Updated information can be found in http://www.latticeaverages.org.

\bibitem{Colangelolat12}
G.~Colangelo, 
  PoS LATTICE {\bf 2012} (2012) 021.

\bibitem{fKoverfpi}
J.~Laiho and R.~S.~Van de Water,
PoS LATTICE {\bf 2011}, 293 (2011)
  [arXiv:1112.4861 [hep-lat]];
S.~D\"urr {\it et al.},
  Phys.\ Rev.\ D {\bf 81}, 054507 (2010)
  [arXiv:1001.4692 [hep-lat]]; 
  C.~Aubin {\it et al.}  [MILC Collaboration],
  Phys.\ Rev.\ D {\bf 70}, 114501 (2004)
  [hep-lat/0407028]; 
A.~Bazavov {\it et al.}  [MILC Collaboration],
  PoS LATTICE {\bf 2010}, 074 (2010)
  [arXiv:1012.0868 [hep-lat]];
Y.~Aoki {\it et al.}  [RBC and UKQCD Collaborations],
  Phys.\ Rev.\ D {\bf 83}, 074508 (2011)
  [arXiv:1011.0892 [hep-lat]];
E.~Follana {\it et al.}  [HPQCD and UKQCD Collaborations],
  Phys.\ Rev.\ Lett.\  {\bf 100}, 062002 (2008)
  [arXiv:0706.1726 [hep-lat]].

\bibitem{Antonelli10}
 M.~Antonelli {\it et al.},
  Eur.\ Phys.\ J.\ C {\bf 69} (2010) 399
  [arXiv:1005.2323 [hep-ph]].


\bibitem{Cirigliano10}
 V.~Cirigliano, J.~Jenkins and M.~Gonzalez-Alonso,
  Nucl.\ Phys.\ B {\bf 830} (2010) 95
  [arXiv:0908.1754 [hep-ph]].

\bibitem{KtopiFNALMILC12}
 A.~Bazavov, {\it et al.},
  arXiv:1212.4993 [hep-lat].


\bibitem{fplus}
P.~A.~Boyle {\it et al.}  [RBC-UKQCD Collaboration],
  Eur.\ Phys.\ J.\ C {\bf 69} (2010) 159
  [arXiv:1004.0886 [hep-lat]]; 
 V.~Lubicz {\it et al.} [ETM Collaboration],
  Phys.\ Rev.\  {\bf D80 } (2009)  111502.
  [arXiv:0906.4728 [hep-lat]];

\bibitem{fKpiproceedings}
T.~Kaneko {\it et al.}  [JLQCD Collaboration], 
  PoS LATTICE {\bf 2012} (2012) 111
  [arXiv:1211.6180 [hep-lat]]; 
E.~G\'amiz {\it et al.}, 
  PoS LATTICE {\bf 2012} (2012) 113
  [arXiv:1211.0751 [hep-lat]]; 
P.~A.~Boyle, J.~M.~Flynn, A.~Juttner, C.~Sachrajda, K.~Sivalingam and J.~M.~Zanotti,
  PoS LATTICE {\bf 2012}, 112 (2012)
  [arXiv:1212.3188 [hep-lat]].


\bibitem{BK2012}
 T.~Bae {\it et al.},
  Phys.\ Rev.\ Lett.\  {\bf 109} (2012) 041601
  [arXiv:1111.5698 [hep-lat]]; 
S.~Durr {\it et al.},
  Phys.\ Lett.\ B {\bf 705} (2011) 477
  [arXiv:1106.3230 [hep-lat]]; 
 J.~Laiho and R.~S.~Van de Water,
  PoS LATTICE {\bf 2011} (2011) 293
  [arXiv:1112.4861 [hep-lat]]; 
  RBC and UKQCD Collaborations, 
  arXiv:1208.4412 [hep-lat].


\bibitem{etacc}
J.~Brod and M.~Gorbahn,
  Phys.\ Rev.\ Lett.\  {\bf 108} (2012) 121801
  [arXiv:1108.2036 [hep-ph]].

\bibitem{Boyle:2012qb}
  P.~A.~Boyle {\it et al.}  [RBC and UKQCD Collaborations],
  Phys.\ Rev.\ D {\bf 86} (2012) 054028
  [arXiv:1206.5737 [hep-lat]].

\bibitem{BKBSM}
 V.~Bertone {\it et al.},
  arXiv:1207.1287 [hep-lat].

\bibitem{fDNf2+1+1}
A.~Bazavov {\it et al.}  [Fermilab Lattice and MILC Collaborations],
  PoS LATTICE {\bf 2012} (2012) 159
  [arXiv:1210.8431 [hep-lat]].

\bibitem{Bazavov:2011aa}
  A.~Bazavov {\it et al.}  [Fermilab Lattice and MILC Collaboration],
  Phys.\ Rev.\ D {\bf 85} (2012) 114506
  [arXiv:1112.3051 [hep-lat]].

\bibitem{Na:2012iu}
  H.~Na, C.~T.~H.~Davies, E.~Follana, G.~P.~Lepage and J.~Shigemitsu,
  Phys.\ Rev.\ D {\bf 86} (2012) 054510
  [arXiv:1206.4936 [hep-lat]].


\bibitem{CLEOfD}
B.~I.~Eisenstein {\it et al.}  [CLEO Collaboration],
  Phys.\ Rev.\ D {\bf 78} (2008) 052003
  [arXiv:0806.2112 [hep-ex]].

\bibitem{RosnerStone2012}
J.~L.~Rosner and S.~Stone,
  arXiv:1201.2401 [hep-ex].

\bibitem{BES2012}
BESIII Collaboration; see, for example, talk given at 
\emph{International Workshop on Theory, Phenomenology and
Experiments in Flavour Physics}.

\bibitem{HPQCDfB1}
C.~McNeile, C.~T.~H.~Davies, E.~Follana, K.~Hornbostel and G.~P.~Lepage [HPQCD Collaboration],
  Phys.\ Rev.\ D {\bf 85} (2012) 031503
  [arXiv:1110.4510 [hep-lat]]; 

\bibitem{HPQCDfB2}
 H.~Na, C.~J.~Monahan, C.~T.~H.~Davies, R.~Horgan, G.~P.~Lepage and 
J.~Shigemitsu [HPQCD Collaboration],
  arXiv:1202.4914 [hep-lat]; 

\bibitem{FMILCfB}
A.~Bazavov {\it et al.}  [Fermilab Lattice and MILC Collaboration],
  Phys.\ Rev.\ D {\bf 85} (2012) 114506
  [arXiv:1112.3051 [hep-lat]].

\bibitem{BaBarBtolnu}
J.~P.~Lees {\it et al.}  [BABAR Collaboration], 
  arXiv:1207.0698 [hep-ex].

\bibitem{BelleBtolnu}
I.~Adachi {\it et al.}  [Belle Collaboration], 
  arXiv:1208.4678 [hep-ex].

\bibitem{BtopiFMILC}
 J.~A.~Bailey {\it et al.}, 
  Phys.\ Rev.\ D {\bf 79} (2009) 054507
  [arXiv:0811.3640 [hep-lat]].

\bibitem{Btopiproc}
 C.~M.~Bouchard, G.~P.~Lepage, C.~J.~Monahan, H.~Na and J.~Shigemitsu,
  PoS LATTICE {\bf 2012} (2012) 118
  [arXiv:1210.6992 [hep-lat]]; 
 F.~Bahr, F.~Bernardoni, A.~Ramos, H.~Simma, R.~Sommer and J.~Bulava,
  PoS LATTICE {\bf 2012} (2012) 110
  [arXiv:1210.3478 [hep-lat]]; 
 T.~Kawanai, R.~S.~Van de Water and O.~Witzel,
  PoS LATTICE {\bf 2012} (2012) 109
  [arXiv:1211.0956 [hep-lat]].

\bibitem{HFAG12}
Y.~Amhis {\it et al.}  [Heavy Flavor Averaging Group Collaboration],
  arXiv:1207.1158 [hep-ex].


\bibitem{BtoDstar}
C.~Bernard {\it et al.},
  Phys.\ Rev.\ D {\bf 79} (2009) 014506
  [arXiv:0808.2519 [hep-lat]].


\bibitem{FMILCBtoD}
S.~-W.~Qiu, C.~DeTar, D.~Du, A.~S.~Kronfeld, J.~Laiho and R.~Van de Water,
  PoS LATTICE {\bf 2012} (2012) 119
  [arXiv:1211.2247 [hep-lat]].

\bibitem{RDFNALMILC}
 J.~A.~Bailey {\it et al.},
 Phys.\ Rev.\ Lett.\  {\bf 109} (2012) 071802
  [arXiv:1206.4992 [hep-ph]].

\bibitem{RDBaBar}
J.~P.~Lees {\it et al.}  [BaBar Collaboration],
  arXiv:1205.5442 [hep-ex].

\bibitem{raredecaysHPQCD}
Z.~Liu, S.~Meinel, A.~Hart, R.~R.~Horgan, E.~H.~Muller and M.~Wingate,
  arXiv:1101.2726 [hep-ph].

\bibitem{raredecaysFMILC}
R.~Zhou {\it et al.}  [Fermilab Lattice and MILC Collaborations],
  PoS LATTICE {\bf 2011} (2011) 298
  [arXiv:1111.0981 [hep-lat]].

\bibitem{HPQCDDtoKpi}
 H.~Na, C.~T.~H.~Davies, E.~Follana, G.~P.~Lepage and J.~Shigemitsu,
  Phys.\ Rev.\ D {\bf 82} (2010) 114506
  [arXiv:1008.4562 [hep-lat]]; 
 H.~Na, C.~T.~H.~Davies, E.~Follana, J.~Koponen, G.~P.~Lepage and J.~Shigemitsu,
  Phys.\ Rev.\ D {\bf 84} (2011) 114505
  [arXiv:1109.1501 [hep-lat]].


\bibitem{FMILCDtoKpi}
C.~Aubin {\it et al.}  [Fermilab Lattice and MILC and HPQCD Collaborations],
  Phys.\ Rev.\ Lett.\  {\bf 94} (2005) 011601
  [hep-ph/0408306].


\bibitem{PDG12}
J.~Beringer {\it et al.}  [Particle Data Group Collaboration],
  Phys.\ Rev.\ D {\bf 86} (2012) 010001.

\bibitem{Koponencharm12}
J.~Koponen, C.~T.~H.~Davies and G.~Donald,
  arXiv:1208.6242 [hep-lat].

\bibitem{UTfits1}
  J.~Laiho, E.~Lunghi, R.~Van De Water,
  PoS {\bf FPCP2010 } (2010)  040
  [arXiv:1102.3917 [hep-ph]];

\bibitem{UTfits1bis}
A.~Lenz {\it et al.},
  Phys.\ Rev.\ D {\bf 83} (2011) 036004
  [arXiv:1008.1593 [hep-ph]].

\bibitem{NiersteCKM12}
A.~Lenz {\it et al.},
  Phys.\ Rev.\ D {\bf 86}, 033008 (2012)
  [arXiv:1203.0238 [hep-ph]].

\bibitem{BmixingHPQCD}
 E.~G\'amiz {\it et al.}  [HPQCD Collaboration],
  Phys.\ Rev.\ D {\bf 80} (2009) 014503
  [arXiv:0902.1815 [hep-lat]].

\bibitem{BmixingFMILC}
A.~Bazavov {\it et al.}, 
  arXiv:1205.7013 [hep-lat].

\bibitem{BmixingRBC}
 C.~Albertus {\it et al.},
  Phys.\ Rev.\ D {\bf 82} (2010) 014505
  [arXiv:1001.2023 [hep-lat]].

\bibitem{Bmixinglatt2011}
C.~M.~Bouchard {\it et al.},
  PoS LATTICE {\bf 2011} (2011) 274
  [arXiv:1112.5642 [hep-lat]].

\bibitem{MixinglatETMC12}
 N.~Carrasco {\it et al.},  
  PoS LATTICE {\bf 2012} (2012) 104
  [arXiv:1211.0568 [hep-lat]].

\bibitem{BurasBmumu}
A.~J.~Buras,
  Phys.\ Lett.\ B {\bf 566} (2003) 115
  [hep-ph/0303060].

\bibitem{BsmumuFleischer}
 K.~De Bruyn, R.~Fleischer, R.~Knegjens, P.~Koppenburg, M.~Merk and N.~Tuning,
  Phys.\ Rev.\ D {\bf 86} (2012) 014027
  [arXiv:1204.1735 [hep-ph]]; 
 K.~De Bruyn, R.~Fleischer, R.~Knegjens, P.~Koppenburg, M.~Merk, A.~Pellegrino and N.~Tuning,
  Phys.\ Rev.\ Lett.\  {\bf 109} (2012) 041801
  [arXiv:1204.1737 [hep-ph]].


\bibitem{LHCbDeltaGamma}
R.~Aaij {\it et al.}  [LHCb Collaboration], LHCb-CONF-2012-002.


\bibitem{BurasGirrbach}
 A.~J.~Buras, J.~Girrbach, D.~Guadagnoli and G.~Isidori,
  Eur.\ Phys.\ J.\ C {\bf 72} (2012) 2172
  [arXiv:1208.0934 [hep-ph]].

\bibitem{LHCb:2012ct}
  RAaij {\it et al.}  [LHCb Collaboration],
   arXiv:1211.2674 [Unknown].

\bibitem{fragmfrac}
J.~A.~Bailey {\it et al.} 
  Phys.\ Rev.\ D {\bf 85} (2012) 114502
   [Erratum-ibid.\ D {\bf 86} (2012) 039904]
  [arXiv:1202.6346 [hep-lat]].


\end{thebibliography}
\end{document}